\documentclass[aps,prl,twocolumn,reprint,showpacs,amsmath,amssymb,superscriptaddress,citeautoscript]{revtex4-1}

\usepackage{graphicx}
\usepackage{epstopdf}
\usepackage{color}

\begin{document}

\title{Large zero-field cooled exchange-bias in bulk Mn$_2$PtGa }


\author{A. K. Nayak}\email{nayak@cpfs.mpg.de}
\affiliation{Max Planck Institute for Chemical Physics of Solids, N\"{o}thnitzer Str. 40, 01187 Dresden, Germany}%

\author{M. Nicklas}\email{nicklas@cpfs.mpg.de}
\affiliation{Max Planck Institute for Chemical Physics of Solids, N\"{o}thnitzer Str. 40, 01187 Dresden, Germany}%

\author{S. Chadov}
\affiliation{Max Planck Institute for Chemical Physics of Solids, N\"{o}thnitzer Str. 40, 01187 Dresden, Germany}%

\author{C. Shekhar}
\affiliation{Max Planck Institute for Chemical Physics of Solids, N\"{o}thnitzer Str. 40, 01187 Dresden, Germany}%

\author{Y. Skourski}
\affiliation{Dresden High Magnetic Field Laboratory (HLD), Helmholtz-Zentrum Dresden-Rossendorf, 01328 Dresden, Germany}%

\author{J. Winterlik}
\affiliation{Institut f\"{u}r Anorganische und Analytische Chemie, Johannes Gutenberg-Universit\"{a}t, 55099 Mainz,
Germany}%

\author{C. Felser}
\affiliation{Max Planck Institute for Chemical Physics of Solids, N\"{o}thnitzer Str. 40, 01187 Dresden, Germany}%
\affiliation{Institut f\"{u}r Anorganische und Analytische Chemie, Johannes Gutenberg-Universit\"{a}t, 55099 Mainz, Germany}%

\date{\today}

\begin{abstract}

We report a large exchange-bias (EB) effect after zero-field cooling the new tetragonal Heusler
compound Mn$_2$PtGa from the paramagnetic state. The first-principle calculation and the magnetic
measurements reveal that Mn$_2$PtGa orders ferrimagnetically with some ferromagnetic (FM) inclusions.
We show that ferrimagnetic (FI) ordering is essential to isothermally induce the exchange anisotropy
needed for the zero-field cooled (ZFC) EB during the virgin magnetization process. The complex
magnetic behavior at low temperatures is characterized by the coexistence of a field induced
irreversible magnetic behavior and a spin-glass-like phase. The field induced irreversibility
originates from an unusual first-order FI to  antiferromagnetic transition, whereas, the spin-glass
like state forms due to the existence of anti-site disorder intrinsic to the material.

\end{abstract}

\pacs{75.60.Ej, 75.30.Gw, 72.15.Jf, 75.50.Cc}
\maketitle

The class of Ni$_2$Mn$_{1+x}$\textit{Z}$_{1-x}$ based Heusler alloys exhibit a structural transition
from a high temperature cubic austenite phase to a low temperature tetragonal/orthorhombic
martensitic phase, whereas, a magnetic ordering transition takes place in the cubic phase. The
existence of a first-order structural transition with strong magneto-structural coupling leads to the
observation of various functional properties \cite{Krenke05b,Kainuma06,Liu12,Pasquale05}. In these
off-stoichiometric Mn rich alloys the extra Mn replaces the {\it Z} atoms. The Mn-Mn exchange
interaction is ferromagnetic (FM) within the regular Mn sublattices while it is antiferromagnetic
(AFM) between the Mn atoms occupying the regular Mn sublattice and {\it Z} sublattice
\cite{Enkovaara03,Aksoy09}. The tetragonal/orthorhombic distortion in the martensitic phase enhances
this AFM interaction due to the decrease in the Mn-Mn distance, which also results in a large degree
of magnetic frustration. Furthermore, many Heusler alloys also display anti-site disorder
\cite{Graf11}. All these factors contribute to a complex magnetic state in the low temperature
regime. The observations of spin-glass behavior \cite{Wang11, Chatterjee09, Nayak11} and the
exchange-bias (EB) phenomenon \cite{Wang11, Khan07} are direct evidence of this complex magnetic
state.

A new class of Mn$_2\mathit{YZ}$ based binary and ternary Heusler compounds stabilize in the
cubic/tetragonal crystal structure with high Curie temperatures ($T_C$). These materials are
attractive candidates for spin-torque transfer devices \cite{Mizukami11,Balke07,Winterlik08,Coey11}.
The magnetic interaction in most of these compounds is found to be ferrimagnetic (FI) in nature
\cite{Balke07,Winterlik11}. The magnetic circular dichroism in x-ray absorption (XMCD) measurements
in Mn$_{3-x}$Co$_x$Ga compounds show direct evidence of the FI ordering \cite{Klaer11}. In a recent
work it is found that Mn$_2$PtIn, which crystallizes in a tetragonal structure with FI ordering,
exhibits an inhomogeneous magnetic state at low temperature  and shows a weak conventional EB effect
\cite{Nayak12}. In this Letter, we report a large unconventional EB effect obtained after zero-field
cooling the Heusler compound Mn$_2$PtGa from its paramagnetic state. To further elucidate the
microscopic origin of the EB behavior we carried out a theoretical investigation on the magnetic
structure. Based on our experimental and theoretical results we propose a phenomenological model for
the large zero-field cooled (ZFC) EB.

Polycrystalline ingots of Mn$_2$PtGa were prepared by arc melting stoichiometric amounts of the
constituent elements and subsequent annealing for one week at 1273\,K. The samples were structurally
characterized by X-ray powder diffraction. The details of the structural analysis is given in the
supplementary material \cite{suppl}. The physical quantities were investigated utilizing Quantum
Design measurement systems. The pulsed magnetic field experiments were performed at the Dresden High
Magnetic Field Laboratory. We have also performed a calculation to map out the magnetic structure
using the PYA-LMTO program package \cite{VWN80}.

\begin{figure}[tb!]
\includegraphics[angle=0,width=8cm,clip]{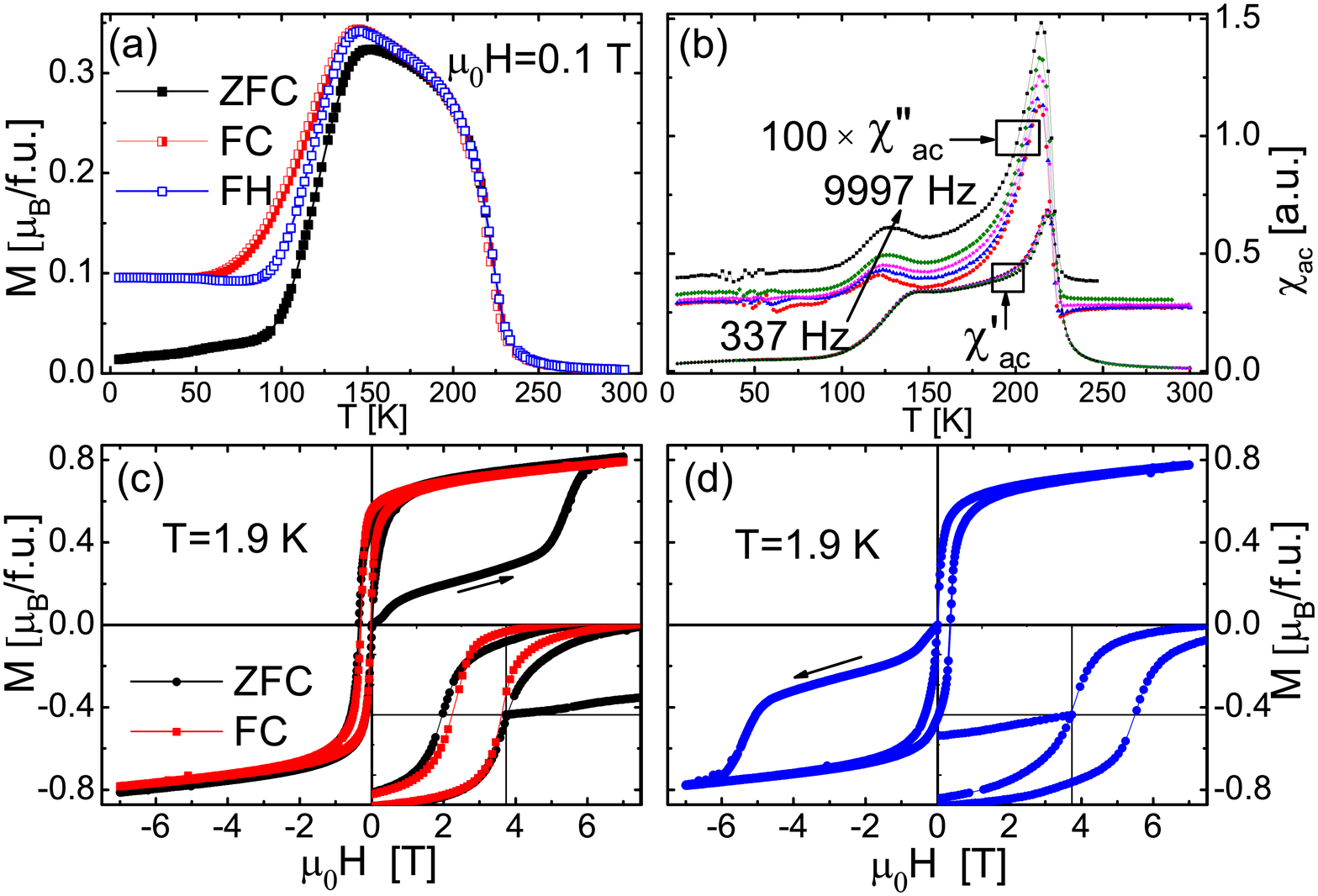}
\caption{\label{FIG1}(Color online) (a) $M(T)$ at 0.1\,T measured in ZFC, FC, and FH cycles. In the
ZFC mode, the sample was initially cooled to 2\,K in 0~T and data were taken upon increasing
temperature in applied field. In the FC mode, data were collected while cooling in field and
subsequently in FH mode data were collected during heating. (b) Real ($\chi'$) and imaginary
($\chi''$) part of the ac-susceptibility measured with different frequencies and an amplitude of
5~Oe. (c) ZFC and FC $M(H)$ loops at 1.9~K performed as $0\rightarrow 7$\,T
$\rightarrow-7$\,T$\rightarrow7$\,T(d) ZFC $M(H)$ loop at 1.9~K performed as $0\rightarrow -7$\,T
$\rightarrow7$\,T $\rightarrow-7$\,T. The insets show a magnified view around $H=0$.}
\end{figure}

Mn$_2$PtGa undergoes a paramagnetic (PM) to FM(FI) transition at $T_C=230$\,K as exemplified in the
magnetization, $M(T)$, and ac-susceptibility, $\chi(T)$, data in Fig.\,\ref{FIG1}a and b. With
further decreasing temperature $M(T)$ exhibits a sudden drop at 150\,K. To probe the nature of this
transition we have measured field cooled (FC) and field heated (FH) $M(T)$ curves that show a
significant hysteresis. The presence of such a thermal hysteresis between the FC and FH curves is a
strong evidence for the first-order nature of a phase transition \cite{Roy04}. Therefore, we argue
that Mn$_2$PtGa undergoes a first-order FM(FI) to AFM transition at 150\,K. However, the existence of
a small irreversibility between ZFC and FC curves suggests that the low temperature phase is not
perfectly AFM in nature. This irreversibility mainly originates from the presence of anti-site
disorder that gives rise to a magnetically inhomogeneous state. We note that the degree of anti-site
disorder can be tuned by the annealing/quenching procedure. The imaginary part of the
ac-susceptibility, $\chi''(T)$ (Fig.\,\ref{FIG1}b), shows a maximum around 120\,K which exhibits a
weak frequency dependence indicating the presence of spin- or cluster-glass like states.

Magnetization loops, $M(H)$, have been measured using a ZFC or FC protocol, as indicated in
Fig.\,\ref{FIG1}c. The most fascinating behavior in the ZFC $M(H)$ data is the shift of the
hysteresis loop in negative field direction by about 0.17~T. To verify this effect we have measured
the $M(H)$ loop also in the opposite direction (Fig.\,\ref{FIG1}d). As expected, this shifts by
0.17\,T to the positive field direction. Therefore, we conclude that the observed EB-like behavior is
intrinsic to Mn$_2$PtGa. Thus, it is possible to induce the exchange anisotropy by zero-field cooling
from the paramagnetic state. The direction of the anisotropy field depends on the initial direction
of the external field. In general, the EB effect in conventional exchange-coupled systems appears
only after field cooling from a temperature above the N\'{e}el temperature of the AFM material. Thus,
when the virgin $M(H)$ curve is measured in positive field direction Mn$_2$PtGa behaves as cooled in
the presence of positive field and vice versa. The $M(H)$ loop taken at 1.9\,K after field cooling in
7\,T shows approximately the same shifting as the ZFC loop (see Fig.\,\ref{FIG1}c). Therefore, in
case of the ZFC EB, scanning the virgin magnetization process provides the roll of the cooling field
required in the conventional EB systems. We note that the ZFC virgin $M(H)$ curve at 1.9\,K displays
a sharp magnetization change at 4.8\,T, indicating a field-induced first-order metamagnetic
transition from an AFM to a FI phase, which will be addressed later.

\begin{figure}[tb!]
\includegraphics[angle=0,width=8cm,clip]{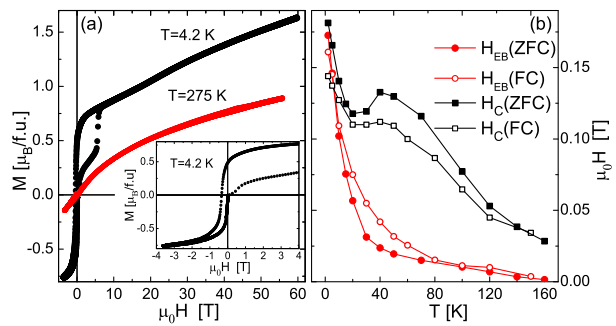}
\caption{\label{FIG2} (Color online) (a) $M(H)$ isotherms at 4.2\,K and 275\,K in fields up to 60\,T.
Inset: data at 4.2\,K in the low field range. (b) $H_{EB}(T)$ and $H_C(T)$ taken from the ZFC (close
symbols) and FC (open symbols) $M(H)$ isotherms.}
\end{figure}

$M(H)$ measured in pulsed fields does not show any saturation up to 60\,T (see Fig.\,\ref{FIG2}a).
The sample shows a magnetization of $0.8\,\mu_B/{\rm f.u.}$ in 7~T at 1.9~K and $1.6\,\mu_B/{\rm
f.u.}$ in 60\,T at 4.2\,K. The small value of the magnetization and the still increasing $M(H)$
indicates the presence of FI ordering in Mn$_2$PtGa. At 275\,K, which is above $T_C$, we find
paramagnetic behavior. Interestingly, the $M(H)$ curve at 4.2\,K displays a similar shift as that
observed in the low field ZFC $M(H)$ data. This evidences that a field of 60\,T is too small to
affect the exchange anisotropy produced in the sample during the initial magnetization process. The
temperature dependence of the EB field ($H_{EB}$) and the coercive field ($H_C$) deduced from the ZFC
and FC hysteresis loops are shown in Fig.\,\ref{FIG2}b. In FC mode the sample was cooled in 7\,T from
300\,K to the desired temperature. $H_{EB}$ and $H_C$ are calculated using $H_{EB} = |H_1 + H_2|/2$
and $H_C = |H_1 - H_2|/2$, where $H_1$ and $H_2$ are the lower and upper cut-off fields. At $1.9$\,K,
$H_{EB}$ possesses a maximum value of around 0.17\,T in ZFC mode and 0.16~T in FC mode. $H_{EB}(T)$
and  $H_C(T)$ exhibit a monotonic decrease upon increasing temperature, which is for $H_C(T)$
interrupted by a local maximum around $40$\,K. The origin of this anomaly will be discussed later.

\begin{figure}[tb!]
\includegraphics[angle=0,width=8cm,clip]{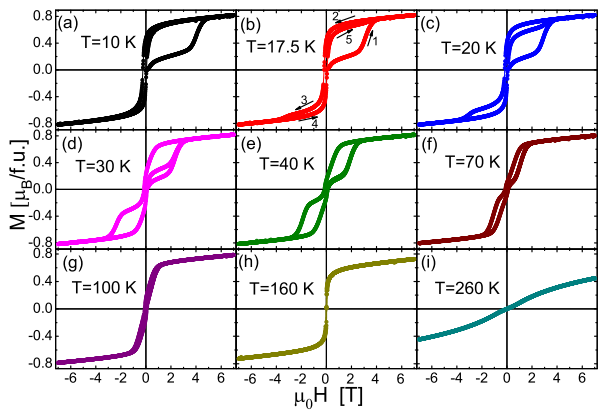}
\caption{\label{FIG3} (Color online) ZFC $M(H)$ loops taken at different $T$.}
\end{figure}

We will now turn our focus to the details of the magnetic phase diagram of Mn$_2$PtGa. Figure
\,\ref{FIG3} displays the ZFC $M(H)$ loops at various temperatures. The 10\,K $M(H)$ loop shows a
similar behavior as that at 1.9\,K (Fig.\,\ref{FIG1}c) with a reduction of the field required to
induce the metamagnetic transition. Interestingly, the metamagnetic transition is only observed in
the virgin curve for $T \leq 15~K$. However, at 17.5\,K the loop shows a signature of the
metamagnetic transition in the negative field path. Upon increasing temperature this signature
becomes more pronounced for negative as well as positive fields. The presence of the virgin curve
outside the envelope loop and the incomplete metamagnetic transition results in an irreversible
hysteresis loop for $T < 40$\,K. The $M(H)$ loops for $T \geq 40$\,K are reversible in both the
quadrants, also the virgin curve coincides with the corresponding part of the hysteresis loop. The
critical field of the metamagnetic transition decreases significantly upon increasing temperature.
The feature of the metamagnetic transition completely vanishes at 160\,K, where we find soft FM
behavior and finally a paramagnetic $M(H)$ loop at 260~K. However, the loops remain shifted in the
negative field direction for $T \leq 160$\,K. From the above observations we can divide the whole
temperature range into different intervals: $0 < T \leq 15$\,K, the AFM phase, which was converted to
FI, cannot be recovered by any number of field cycling. This results in a large field induced
irreversibility in the $M(H)$ loops. In the temperature range $15 {\rm\,K} < T < 40$\,K the AFM phase
can be partially restored. At higher temperatures ($40{\rm\,K} < T < 150$\,K), the AFM phase that was
initially converted to the FI phase is recovered fully to its initial state by cycling the magnetic
field. Here the hysteresis loops are fully reversible and the AFM phase returns to its initial state
by reducing the field to zero. The peak in $H_C(T)$ around $40$\,K, shown in Fig.\,\ref{FIG2}b,
originates from the irreversible hysteresis loops in the range of $17.5 {\rm\,K} \leq T \leq 40$\,K.
This results in a nominal increase of $H_C(T)$ upon increasing temperature for $17.5 {\rm\,K} \leq T
\leq 40$\,K followed by a reduction for reversible loops at higher temperatures.

The disorder influenced first-order magnetic to magnetic transition in Mn$_2$PtGa is found for the
first time in any Heusler compound. The observation of different magnitudes of field-induced
irreversibility is a direct consequence of the competition between the potential and thermal energy.
Though the AFM phase possesses a lower energy at $H=0$ for $T\leq15$\,K, the field induced FI phase
forms a pinning potential, where disorder provides the pinning centers. When the field is reduced to
zero the thermal energy cannot overcome this potential and the AFM phase is not restored resulting in
the observed irreversibilities. At higher temperatures the thermal energy becomes large enough to
allow for a partial recovery of the AFM phase by field cycling. Only above 40\,K the thermal energy
becomes larger than the pinning potential and a reversible hysteresis loop is observed. Similar
phenomena have been reported in systems undergoing a first-order magnetic to magnetic transition
\cite{Sengupta06,Dho03}. Although some of the Heusler systems with martensitic transition show a
similar type of field induced irreversibility, the underlying physics for such a behavior
categorically differs from the case of Mn$_2$PtGa. In the former materials the field induced
structural change plays a major role in inducing the irreversibilities, which become stronger on
approaching the martensitic transition upon increasing temperature \cite{Sharma07,Nayak10}. This is
in contrast to Mn$_2$PtGa which does not show any structural transition.


\begin{figure}[tb!]
\includegraphics[angle=0,width=8cm,clip]{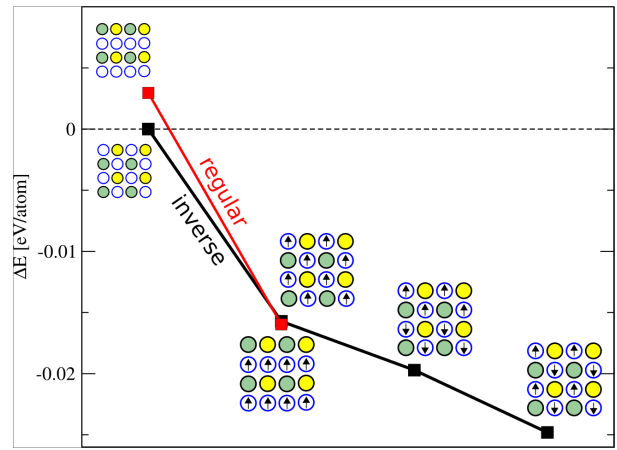}
\caption{\label{FIG4} (Color online) Total energies for various magnetic configurations in case of
  inverse (black) and regular (red line) Heusler structures. The energy
of the inverse Heusler structure in the nonmagnetic state is taken as
reference. The chemical composition of each configuration is
schematically depicted by the 4$\times$4-atom diagrams drawn closer to the
corresponding energy points. Light green, yellow and hollow spheres mark
Pt, Ga and Mn atoms, respectively. Arrows on Mn spheres indicate the
directions of the local magnetic moments.}
\end{figure}

To map out the exact magnetic configuration in Mn$_2$PtGa, we have performed calculations using the
fully-relativistic linearized muffin-tin orbitals band-structure method within the PYA-LMTO program
package \cite{VWN80}. The unit cell parameters were taken from experiment. The exchange-correlation
potential was treated using the Vosko-Wilk-Nusair form of the local density approximation \cite{PYA}.
The comparison of the total energy calculated for the different configurations (see Fig.\,\ref{FIG4})
indicates that the nonmagnetic configurations are energetically the most unfavorable (regular and
inverse variants). The energy is substantially reduced by switching on the magnetism on the Mn atoms.
For a ferromagnetic configuration this results into a large magnetic moment: 7.2~$\mu_{\rm B}$/f.u.\
in case of the regular and 7.12~$\mu_{\rm B}$/f.u.\ for the inverse Heusler structure. In this fully
ferromagnetic setup both regular and inverse structures possess rather similar total energies of
about -15~meV/atom compared with the nonmagnetic inverse structure. For the regular Heusler variant
the ferromagnetic state is the most stable. Enforcing any type of antiparallel alignment of Mn
moments within the regular structure always leads to a locally nonmagnetic state. On the other hand,
by varying the magnetic configurations further within the inverse structure, the total energy still
can be lowered. By reverting the directions of the magnetic moments in each second pair of layers, we
arrive in the fully compensated antiferromagnetic state, with an energy lowered by about 4~meV/atom
compared with the ferromagnetic state. The lowest energy configuration is achieved when the Mn moment
is reverted within each layer (FI state). Since the positions of the Mn atoms in the adjacent layers
within the inverted Heusler structure are nonequivalent by symmetry, the total magnetic moment is not
completely compensated: a small total moment of about 0.55~$\mu_{\rm B}$/f.u., which is the result of
3.65~$\mu_{\rm B}$/f.u.\,[Mn(I)] - 3.1~$\mu_{\rm B}$/f.u.\,[Mn(II)], remains. Our results emphasize
that the formation of the magnetic state is driven by two mechanisms -- ferromagnetic exchange
between Mn atoms at long and antiferromagnetic at short distances.

The observation of a magnetization of $0.8\,\mu_B/{\rm f.u.}$ at 7~T is slightly larger than the
calculated value (0.55~$\mu_{\rm B}$/f.u.). This can be explained by ferromagnetic clusters embedded
in the strongly-compensated ferrimagnetic host. The ferromagnetic ordering inside the clusters is due
to their regular Heusler structure, i.\,e. the interchange between Mn and Pt atoms (the so-called
anti-site disorder, which often takes place in Heusler materials). Samples prepared under different
annealing and/or quenching conditions exhibit a slightly changed magnetization. This is most likely
connected to an atomic rearrangement toward the more stable, inverse Heusler structure. The latter
phase forms an almost compensated host, which surrounds the ferromagnetic clusters.

Based on the first-principle calculations and experimental results we propose a simple model for the
ZFC EB in Mn$_2$PtGa. We assume an exchange interaction between two dissimilar magnetically
anisotropic phases. The irreversibility between ZFC and FC $M(T)$ curves in low fields ($\leq
0.1$\,T), the frequency dependence of $\chi''(T)$ below $T_C$, and the theoretical calculations
indicate the presence of magnetically inhomogeneous states originating from local FM clusters in a FI
background. Upon applying field the virgin magnetization process will try to align the moments in the
FI and FM phases along the field direction, which implies that the interface spins of the FI and FM
phases will align in the same direction to minimize the energy. This sets up the exchange interaction
between the magnetically soft FM phase and the magnetically hard FI phase. Further, this implies an
increase of the coercive field in a perfectly ordered material, because the ferrimagnetically ordered
moments collectively change their direction \cite{Nogues05}. However, one would expect a rough
interface with disorder and uncompensated moments in the FI phase and additional defects in the bulk
\cite{Nogues05,Nogues99,Kiwi01}. All of these factors collectively contribute to the ZFC EB behavior.
The fact that $H_{EB}$ obtained from the FC $M(H)$ loops measured after field cooling in 7\,T, where
the sample is only in the FI phase below $T_C$, and the $H_{EB}$ derived from the ZFC loops almost
match rule out any significance of the first-order FI to AFM transition in the formation of the ZFC
EB. This is furthermore supported by the observation of a small EB at 160~K, which is above the FI to
AFM transition. The pulsed field experiments up to 60\,T find only a very small change in $H_{EB}$.
This shows that once the exchange interaction is set up the field does not change the FI phase
significantly to observe a change in $H_{EB}$. These findings distinguish Mn$_2$PtGa from the
off-stoichiometric Ni-Mn-In Heusler alloys where recently a ZFC EB has been reported \cite{Wang11}.
Mn$_2$PtGa exhibits a too weak frequency dependence of the peak in $\chi''(T)$ to indicate super
spin-glass behavior resulting in super FM clusters during the initial magnetization process as in
Ni-Mn-In \cite{Wang11}. One common factor, however, is the existence of FI ordering in both systems.
Therefore, we argue that the presence of FI ordering with embedded FM clusters is an important
ingredient for the appearance of a ZFC EB.

In conclusion, we have synthesized and studied the new Heusler compound Mn$_2$PtGa. Though,
Mn$_2$PtGa orders ferrimagnetically at $T_C=230$\,K, it undergoes an unusual first order FI to AFM
phase transition below $T_C$. We demonstrated, to our knowledge for the first time, the presence of a
ZFC EB effect in a stoichiometric bulk Heusler compound. We further show that the appearance of
this ZFC EB effect is related to the presence of FI ordering with embedded FM clusters. We ruled out
any role of the first-order transition in the observation of the ZFC EB.

We thank J.\,A.\,Mydosh for valuable discussions on the present work. This work was financially
supported by the Deutsche Forschungsgemeinschaft DFG (Project No.\,TP 2.3-A of Research Unit FOR 1464
ASPIMATT) and by the ERC Advanced Grant (291472) "Idea Heusler". The experiments at the High Magnetic
Field Laboratory Dresden (HLD) were sponsored by Euro-MagNET II under the European Union contract
228043.

\end{document}